# Minimalistic Terminal Editor for Julia Programming - MinTEJ: A Friendly Approach for a Scientific Programmer

Poornachandratejasvi Laxman Bhattar[1], Payal V. Dahiwale[2], Krishnarjunulu Thota[3], and Anurag Sharma[4]
Hitachi Energy Technology Services Private Limited, Bangalore, India [1,3], Indian Institute of Technology Bombay, Mumbai, India[2], and Newcastle University in Singapore, Singapore[4]
poornachandratejasvi.bhattar@hitachienergy.com[1]

***Abstract* — Developers rely on lightweight, terminal-centric workflows for rapid code iteration. However, within a unified environment for Julia programming language, existing tools provide limited support for integrated workflow such as editing, execution, file management, and debugging. As a result, developers frequently incur context-switching overhead and fragmented tool interactions. Therefore, the proposed work predominantly focuses on the minimalistic approach for developing native terminal editor for Julia programming language. This paper introduces MinTEJ, a terminal-based editor built in Julia, and proposes a Sequential Modal Interaction Architecture (SMIA) that unifies file management, code editing, execution, and debugging through a command-oriented workflow. The presented work formalizes model interaction and reduces cognitive load & errors while transitioning among different modes. In SMIA, buffer is the central data structure that persists across all modes. Each mode interprets and manipulates the buffer according to mode-specific rules. The central controller mediates access to the buffer and enforces sequential transitions between modes. To evaluate the approach, the performance benchmarking of MinTEJ is compared against existing tools i.e., VS code and Notepad++. The effectiveness of the proposed MinTEJ is evaluated based on memory consumption and CPU utilization demonstrating that it has less resource overhead. Findings suggest that integrated terminal-based editor environment is a practical lightweight software tool enabling efficient iterative development.**

## I. INTRODUCTION

Programming languages are fascinating in the world of scientific computing and their application in the general programming. Code editors developed for these programming languages assist developers to process, draft, manage, and maintain the codes [1]. Over the period of time, there have been inception of many programming languages – some of the classic programming languages are LISP, FORTRAN, C, C++, Python, and Java. Editors provide the flexibility to plug-in programming languages for easy development. There are class of editor such as an integrated development environment for writing and debugging the codes written in one or more programming languages [2]. Recently, Julia is an emerging scientific computing language and it is gaining popularity because of its simplicity like python and execution speed similar to C language [3], [4], [5], [6]. It is built on low level virtual machine (LLVM) and Just in time (JIT) compiler [7], [8]. There are limited editors and integrated development environments (IED's) that can be used for Julia programming, but there are no Julia-specific IEDs at present [9]. The popular IDE such as Juno and VS code provide flexible development of Julia but in recent times, Juno development has been limited to only maintenance and VS-Code is actively used for Julia development [9]. Resource intensive IDE's affects the machine performance. There are lighter version of editors such as Vi, Notepad++, sublime text and many more which are used to edit codes [10]. These are actively used to write and edit the code but they are limited in capabilities such as debugging and code execution.

Julia is relatively new language, where existing IDEs are evolving to support it. The potential issues highlighted by the user community are mentioned in Julia Discourse with the thread as "Extremely slow debugging in VS Code" which indicates VS Code is slow to load the Julia [11]. In addition, debugging takes plenty amount of time and slows the process. The potential other issues reported in the Discourse of Julia with the thread such as "VScode + Julia, slow" and similar issues, are mentioned in the GitHub as "slower response of VS code editor" [12], [13]. Similar issues are reported in reddit forum with the thread "VsCode Experience" [14]. The potential cause of such issues in code IDEs may include multiple reasons, such as machine specification, resource-intensive application, and unexpected behaviour of programs [15], [16]. The lightweight editors mentioned earlier are well-suited for rapid code changes and their simplicity to manage the code with less overhead. However, these editors are more specifically for text editing and understanding the code flow, whereas the real-interaction and code introspection are challenging unlike the IDEs. In notepad++, the Julia text editing is supported by the external XML script. At present, the read, evaluate, and print loop (REPL) are not readily available in these text editors and the workflow execution relies on AutoHotkeys. The configuration to plugin in Vi editor is cumbersome for the beginners. The text editors do not facilitate real-time interaction with code using REPL or live code introspections.

Editors are written in various programming languages such as TypeScript, C, C++, LISP and PERL, etc. and rely on the language server's plugin for the programming languages to be used. Often plug-in may lead to the overhead or intensive resource usage. Code editors may not have access to Julia AST or Julia code intermediate language which makes debugging the code more challenging. At present, the available text editors, and IDEs should upgrade to increase the flexibility of Julia usage [17]. However, in Julia programming external packages used for code tracking, debugging, and code introspection are done using infiltrator.jl, and debugger.jl. The native Julia code editor will be efficient with the less resource requirements as compared to non-native code packages.

To summarize, the existing Julia IDEs and editors have many challenges, such as the response speed of IDEs for code writing and execution, the resource intensiveness, lack of REPL orchestration, and the accessibility for the intermediate script and AST. To overcome the above-mentioned challenges, this work proposes the first native Julia editor and the programming package - MinTEJ (minimalistic terminal editor for Julia programming) for code writing, management, and execution, as well as for code introspection. The major contributions of the proposed work and detailed work packages are mentioned below:

1. *Sequential modal interaction architecture (SMIA):* This paper introduces a formal interaction model that has well-defined architecture governed by a central mode. SMIA provides clear semantics for mode transitions and with structured workflows in terminal-centric environment named MinTEJ. A unified buffer abstraction persists across all modes, enabling consistent state representation for editing, file management, execution, and debugging. The buffer serves as the core data structure of SMIA.
2. *Option rich julia editor environment:* The proposed MinTEJ editor has four operation modes to cover the necessary operations of an editor, which include code management and execution, file handling, debugging, and a quick information guide. These modes of operation are described as Editor mode (edm) – to write, delete, save the code, and other necessary commands. File management mode (fms) – navigate the file, save, rename, and remove and many more operations. Execution mode (exe) – code execution and REPL like mode. Debug mode (db) which provide the stepwise running of code and applying the breakpoint based on the AST of the code. Syntax mode (syntx) is a knowledge base where the user can look for command syntax and existing examples.
3. *Minimalistic approach:* MinTEJ has abstract syntax-based debugger that provides the provision of live code introspection, REPL orchestration, and overplay relays which makes it convenient for users. MinTEJ is a native and Julia-specific editor that can adapt easily to Julia updates and modifications. The developed editor does not rely on the resource-intensive package and can be customized as per the requirements which grants it more freedom and reliability to use the developed package.

The working examples are provided to demonstrate the effectiveness of the developed package and the resource comparison of the proposed work is included to understand the application behaviour. In the paper, methodology and algorithms are provided in the Section II, Section III consist of results and discussion, and Section IV includes the future changes and conclusion.

## II. Detailed Framework and Algorithm

The functionalities of the proposed work are specifically designed considering Julia-eccentric users to provide clear, transparent, and modular workflow approach. The terminal applications provide the interactive Julia environment in the designed package. Sequential Modal Interaction Architecture (SMIA) of proposed MinTEJ is given in Figure 1. The developed work package in Julia assists the learners with a new approach to interact and introspect the code. This workflow of the proposed minimalistic terminal editor is inspired by the teaching paradigm which provides the users a flexibility to interact and manage the code, and provides file management, code execution, and REPL like environment in the developed Julia package– MinTEJ. The detailed framework is represented in Figure 2 that shows the organization of the proposed work-package demonstrating its modular approach for flexible control and operation in Julia programming. The following section discusses the design process and detailed algorithm for each mode of operation in MinTEJ terminal editor.

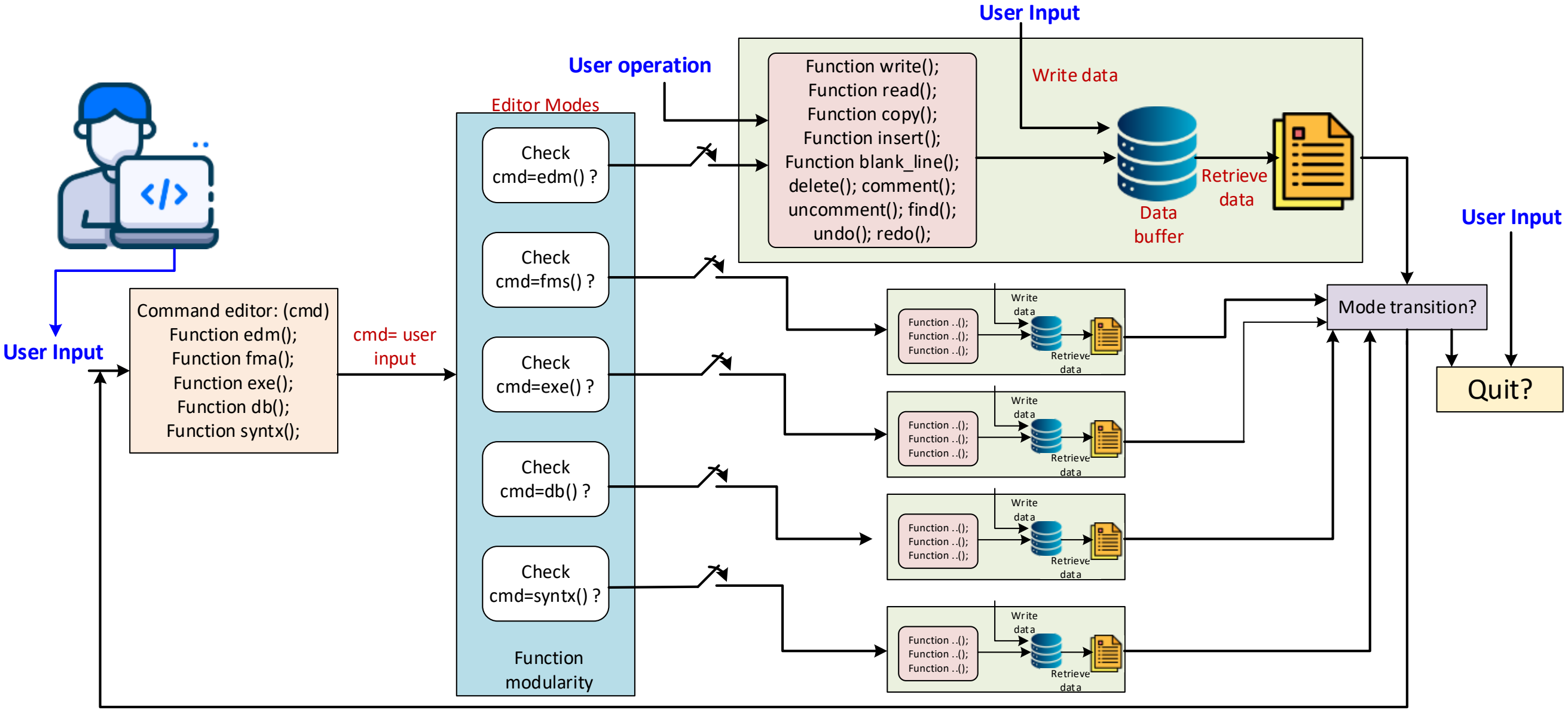

Figure 1. Sequential modal interaction architecture (SMIA) of the proposed MinTEJ

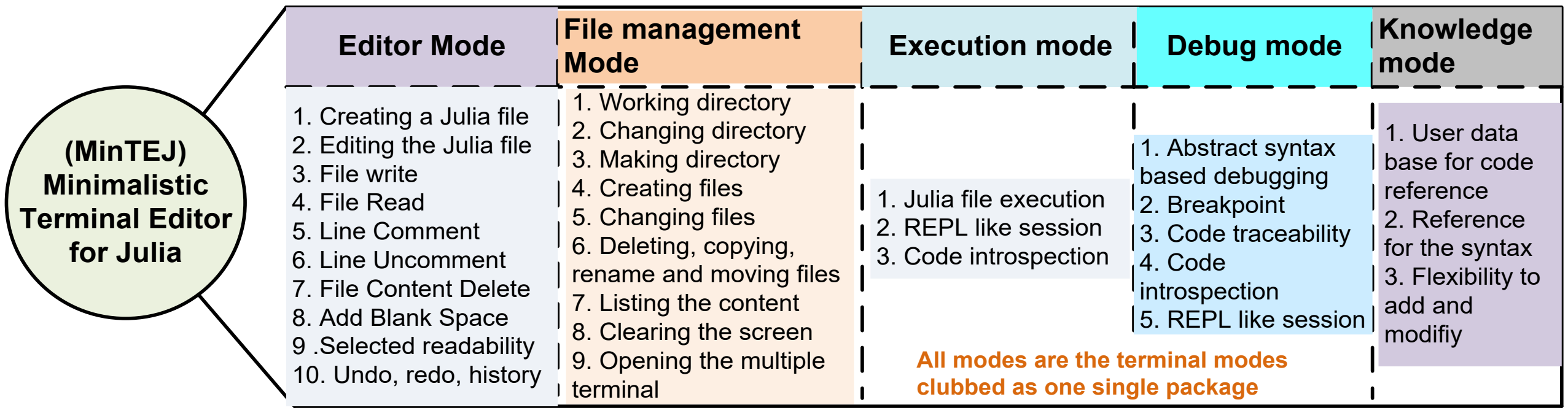

Figure 2. Sequential modal interaction architecture (SMIA) of minimalistic terminal editor for Julia (MinTEJ) with the detailed functionality of operating mode and its corresponding functions.

### A. *Editor mode*

The editor mode in the MinTEJ adopts modal interactive paradigm for editing and introspection of code. The adopted paradigm helps to maintain clarity and coordination among each module with efficient navigation mechanism, editing, execution, and debugging of code. The simplified structure helps to reduce the cognitive load while working upon the editor. The generic layout of the editor mode includes the buffer mechanism. The sequential buffer data structure includes quintuple of actual data, buffer capacity, index sets, and head & tail pointers as given in (1). The buffer state space is defined as in (2), that shows the condition of buffer in terms of $H, T, C$.

$$\mathbb{B} =< \mathrm{A, C, I, H, T} > \\ \mathrm{I} = \{0,1, \dots, \mathrm{C}-1\}, \mathrm{H, T} = \{0,1, \dots, \mathrm{C}\} \tag{1}$$

$$\mathrm{S} = \{(\mathrm{H, T}) \in 0 < \mathrm{H} < \mathrm{T} < \mathrm{C}\} \tag{2}$$

The current number of data stored in the buffer is determined using $X \rightarrow (T - H)$. The data store in the buffer is given in (3). The partial state transition while writing the data into the buffer follows $W: S \times D \rightharpoonup S$, where $\rightharpoonup$ denotes the partial function, $D$ is data set and the transition is $A(T) \coloneqq d, d \in D$. While writing in the buffer the tail pointer shifts to the next index while head pointer remains same as shown in writing rule given in (3). Similarly, the reading operation has the partial transaction as $R: S \rightharpoonup S \times D$, and $d \coloneqq A(H)$. In the data retrieval rule, the head pointer increases by 1 index and the tail pointer remains same as given in (4).

$$\mathbb{B}_{val} = \{A_i |\ i \in [H, T-1]\} \tag{3}$$

$$(H, T) \xrightarrow{write\ (d)} (H, T+1) \tag{4}$$

$$(H, T) \xrightarrow{read\ (d)} (H+1, T) \tag{5}$$

In this mode, the user inputs are converted into the standard string format inputs to the system and these strings are stored in the vectorized format. This mechanism can be seen as array of strings and the array operations are used to modify the code. The Algorithm#1 provides the application of buffer logic to write the files during editor mode and other various operations that are triggered in the proposed terminal editor. In the editor mode function "edm" requires the input as filename which is of string type followed by option such as write, save, read, delete and so on for controlling the editor mode. The code structure provided in the editor architecture is an imperative programming that provides more control for users to modularize the code as per the requirement. Option in the edm function demonstrates algorithm modularity where future code extension and modifications are easy for maintaining the code clarity with the modular approach. Options in the edm mode are backed by the print and user input statement for easy usage of the developed editor function. As mentioned, the core functionalities are controlled by buffer logics which are like array manipulation of strings. Consider a sample text stored in Buffer as buffer = ["The sky is blue.", "Sun rises in East.", …, …]. Let say user needs to write after the line "Sun rises in East", the logic is accessing the buffer with index 3 and write the required string into the buffer. Rules like string catenation, deleting, and inserting are applicable in buffer manipulation. Once, the buffer is edited; the program reads the buffer and write the content in file with the desired extension.

*Algorthm#1: Editor mode functionalities and pseudo-code*

```
1.  Function edm(filename, input option)
2.  If input option = write Then
3.     Check file exists? If not Then create one
4.     Create Buffer to hold the vector strings
5.     Open the Standard Input to Read user input into a Buffer
6.  If input = Save Then write the text file and save
7.  Else If input option = read Then
8.       Read file into Buffer
9.       Print content of Buffer
10. Else If input option = delete Then
11.       Prompt user option for line number to delete
12.       Delete the content of Buffer index
13. Else If input option = copy Then
14.       Prompt user option for line number to copy
15.       Copy the content of Buffer index
```

| | |
|---|---|
| 16. | **Else If** input option = comment **Then** |
| 17. | **Prompt** user option for line number to comment |
| 18. | **Read** the content of Buffer index |
| 19. | **Catenate** the string "#" and content of Buffer |
| 20. | **Else If** input option = uncomment **Then** |
| 21. | **Prompt** user option for line number to Uncomment |
| 22. | **Read** the content of Buffer index |
| 23. | **Remove** the string "#" from the content of Buffer |
| 24. | **End** |

In the proposed editor, the files are stored with the "*.jl"* extension. The editor mode stores all the operations performed on the file by the user as history. The *edm* has options such as copy_file to copy content of one file into other, find option to search with keyword and undo & redo option to recover the file based on the previous operations. The developed *edm* functions are called independently in the proposed editor. The *edm* mode is gracefully exited using the keyboard interrupt and editor program hover in the main mode MinTEJ for navigating into another mode.

### *B. File management mode*

The developed package has a functionality to manage and navigate files during the code development and help to store & organize them more effectively. The segmented approach for managing and reorganization of files and directory is considered. In the existing Julia terminal, the file management functionalities are available but they are distributed vaguely which makes it difficult for the user to use the terminal functions. Therefore, the proposed editor provides functionalities such as listing the files, directory, change the path, print current directory, viewing directory structure, renaming, deleting, and copying the files from the directory structures, accessing the multiple terminals, file comparison based upon the size and characters and many more. In addition, the clearing functionality has been incorporated to clear the working window for user readability. File management mode in the minTEJ editor is acronymic as "fms" and the application can hover in the multiple mode using the keyword such as fms, edm, db, and so on. The pseudo-code for file management system is provided in algorithm#2. The different options in the fms can be navigated using the keywords. The proposed fms mode is rich in functionality and modularity in addition to general functionality like file management creation, listing, renaming, and copying, the fms ecosystem also encourages file comparison. The philosophy of comparison between the file involves the conversion of files into different buffers. Each buffer contents are spliced as words and white spaces. The matching between the content is carried using string comparison to ensure the match is found else the difference is shown on the console to the user. Using this functionality, the user can compare multiple code file with .jl extension to identify the mismatch between them. The fms ecosystem also includes the functionality of opening multiple terminals and running the Julia command independently. This eco-system also provides the safety for critical operations such as file deletion using user confirmation for the action. The details of the various operations included in the fms eco-system are presented in Algorithm#2.

*Algorithm#2: Detailed functionalities of file management system and pseudo-code*

| | |
|---|---|
| 1. | **Function fms**(input) |
| 2. | **If** input = list current directory **Then** |
| 3. | Print → working directory |
| 4. | **Else if** input = list content of directory **Then** |
| 5. | **Walk** every content of Directory |
| 6. | Print → Each Content |
| 7. | **Else if** input = Change directory **Then** |
| 8. | **Prompt** the user specified path |
| 9. | **Change** the working path |
| 10. | **Else if** input = list all files **Then** |
| 11. | Print → All files |
| 12. | **Else if** input = Tree **Then** |
| 13. | **For each** root, each directory, each file |
| 14. | Print → root |
| 15. | Print → File |
| 16. | Prints Tree structure |
| 17. | **Else if** input = Clear **Then** |
| 18. | Clear the console |
| 19. | **Else if** input = Delete file **Then** |
| 20. | **Prompt** user confirmation |
| 21. | **Delete** file |
| 22. | **Else if** input = Delete folder **Then** |
| 23. | **Prompt** user confirmation |
| 24. | **Delete** file |
| 25. | **Else if** input = rename **Then** |
| 26. | **Rename** directory/file |
| 27. | **Else if** input = copy **Then** |
| 28. | **Copy** directory/file |
| 29. | **Else if** input = Make directory **Then** |
| 30. | **Create** new directory |
| 31. | **Else if** input = compare **Then** |
| 32. | **Compare** files character to character |
| 33. | **Displays** number of lines |
| 34. | **Displays** size of files |
| 35. | **Else if** input = read **Then** |
| 36. | **Read** out the files |
| 37. | **Displays** content of file |
| 38. | **Else if** input = New terminal **Then** |
| 39. | **Opens** new terminal |
| 40. | **Else** input = info |
| 41. | **Displays** all the command for the user |
| 42. | **End** |

### *C. Execution mode*

Unlike the other available IDEs, the developed terminal editor Julia serves as minimalistic approach for programme execution. Scope of this section is to provide a Julia program execution capability. The MinTEJ is designed to involve the functionality of program execution using the existing Julia command and make it flexible to reuse in editor for code execution which will help the user to quickly edit and run. In this execution mode "*exe*", the code is executed using the command *include("file_name.jl")*. The standard output and error logs are captured by running the *include* command. The *file_name.jl* represents the Julia file that user wants to run.

Running the proposed package isolates the access to the terminal of Julia. Therefore, using the REPL environment directly is not possible. To address this issue, the REPL session is created inside the program to emulate the REPL environment. The principle to utilize the REPL like session expands the philosophy of Julia code evaluation. In this mode, the code is converted to strings and parsed as expressions. These expressions are evaluated inside Julia as a separate code providing the user interaction of REPL like session.

Julia environment is rich in meta programming, and the functionalities such as *Meta.parse()* and *eval* comes much handy to provide the REPL like session in the proposed editor that makes it more suitable to interact with terminal while using the proposed MinTEJ package. The detailed algorithm for the execution mode is provided in Algorithm#3. The presented execution mode provides options such as REPL session and direct code execution. In REPL session, user has option to load the Julia or use the presented terminal to run the own code or quick code. When a user loads the files, the code is read into buffers. Then, the implemented buffer logic converts each line of code into string and prepares the string code using a join to further process as an expression. Similarly, the user should provide standard input and similar logic is utilized to run the code as expression. The REPL like session provides the user to view the environment variables. This functionality acts as *watch* on the variables that user needs to track during the code evaluation.

*Algorithm#3: Pseudo-code for execution module*

```
1.  Function exe(input)
2.  If input = REPL Then
3.     Prompt the user to select mode
4.     If input = file mode Then
5.        Prompt user file name
6.        Read file in buffer
7.     Else
8.        Prompt user to write
9.        Read into buffer
10.    End
11.       Join the buffer
12.       Convert into expression
13.       Evaluate the expression
14.       Prompt user variables to watch
15.       Get the environment variables
16.        Print → Variables
17. Else if input = execution Then
18.       Prompt user input file name
19.       Include("file.jl")
20.       Capture Standard output
21.       Capture error
22.       Print → Output Else Error
23. End
```

## D. Debug mode

The MinTEJ package has a debugging mode and this mode is operated with the user input. This proposed minimalist debugger works based on an abstract syntax tree. In the first step, the input code is converted into Julia expression. For every Julia expression, the AST is traced recursively. The auto step-in function is developed considering the AST. In paradigm of Julia, obtaining the AST using the meta programming is simpler approach using *Meta.parse()*. The AST is traced back recursively and the evaluation of expression is made. The detailed algorithm is provided in Algorithm #4 that provides the recursive collection of AST and evaluation of expression which is used as debugger. AST for a given expression consists of *head* and *args*. The *head* is a symbol or a type, for example, *head* may include the various programming keywords such as *function, for, while if, quote, symbols* and so on. Each *args* consist of body which determines the user functionality and *args*. The *args* (also referred as arguments) in Julia can be obtained using *args[i],* where, *i* is an index. This debugger is purely based on AST traversal and recursive AST ingredients collection that are nodes and edges. The meta programming is used to evaluate the expression and every expression collected recursively is evaluated. Here, the control over the code is maintained using the *readline()* command which pause to seek the user inputs. As seen from the algorithm function in Algorithm #4, step is wrapped as macro, and it is injected in the code. The code expressions are identified based on their *heads* and evaluated as shown in the algorithm# 4. During this process, the given input code is converted to expression, and further expression is broken down as *heads* such as *for, while, if, else, function*, and so on. The algorithm matches the *head* of given expression and loop body and conditions are extracted from AST. Meta expressions are used to evaluate the expression and capture the behaviour of the variables. The code tracks all the input variables specified by user giving an opportunity to introspect the code. The output is printed as the standard output and the complete code is wrapped in the quote expression furthermore, error in this discussed module is handled by try and catch. The code is automatically wrapped into function to debug, and the copy of debug file is created for user reference.

*Algotihm#4: AST based debugger and pseudo-code for the debugging Julia programme based on the code expression*

```
1.  Function step(expr, depth=1)
2.  If expr isa Expression(Expr)
3.      If expression head = :block Then
4.        Return Expr(:block, call step(expr, depth+1) for every expression in args)
5.      If expression head = :if Then
6.        condition= call steps(expression argument[1], depth+1)
7.        then_branch = call steps(expression argument[2], depth+1)
8.        For else_branch
9.         check length(expression argument) == 3
10.        call steps(expression argument[3], depth+1)
11.        Return Expr(:if, condition, then_branch, else_branch)
12.     Else if expression head == :while Then
13.         condition = steps(expression argument[1], depth+1)
14.         body = steps(expression argument[2], depth+1)
15.         Return Expr(:while, condition, body)
16. Else if expression head = :for Then
```

```
17.        loop_header = expression argument[1]
18.        loop_var = loop_header's argument[1]
19.        loop_iter = loop_header's argument[2]
20.        loop_body = steps(expression argument[2], depth+1)
21.        Return quote
22.            Pause user to inspect
23.            Prompt user environment variables
24.            User chooses list of environment variables
25.            Print → variables for inspection
26.        End
27. Else if expression head = :function Then
28.        Repeat steps 1-26 for code introspection
29.        Pause for user inspection
30. Else
31.     Evaluates the expression
32.     Pause of user to introspect
33.     Prompt the user to view variables
34. End
```

Within debugging mode, the macro for breakpoints and variable inspections are developed in the proposed package. In this mode, the user has flexibility to load the breakpoints for a given line number. The algorithm for breakpoint is provided in algorithm #5. The presented debugging algorithm #4 and #5 are combined as a single debug mode and operates as different options such as interactive break point mode. In this mode, the user can inject the macros at the desired location and analyse the code behaviour before the breakpoint. This interactive mode is a REPL session based on the buffer logic and code evaluation. Similarly, user can run the code step-wise using the macro *step*. The macro step is based on the AST and user can interact with the code. Debug mode has provision that user can directly provide the Julia file name as input for debugging or can use the option in REPL like session. Like other sub-packages or modes in the developed package, the debugging code is also modular and has flexibility to modify as per the requirement.

*Algorithm#5: Pseudo-code for break point evaluation, provided as macro*

```
1. Macro breakpoint()
2.     Return quote
3.     Extract the variables in scope
4.     For each variable
5.     Print → each variable
6.     End
7.     Pause()
8.     End
9. End
```

### E. Syntax mode

Syntax mode is user database to load the syntax of the commands with examples. This mode is also known for quick information guide to users. The different command syntax is given in "syntx" mode by providing the input as command name, for example, while, for. This mode is included to provide quick command syntax to users and to encourage the user to freely navigate the syntax without any barrier. The syntax mode covers the brief topics on variables, control statements, functions, array, and print statements. The basic pseudo-code for the syntax mode follows the identifier provided in Algorithm#6.

*Algorithm#6: Pseudo-code for syntax highlighting*

```
1.  Function syntx(input)
2.  If input = for Then
3.  Code = " for k =1:10
4.          println("Index is:, k)
5.          end"
6.  println → code
7.  Else if input = While Then
8.  Code = "global x = 0
9.          While x <=5
10.          x = x + 1
11.          println(x)
12.         end"
13. println(Code)
14. Else if
15.      Add identifier
16.      Define and add the database
17. End
```

## III. Results and Discussion

In this work, the proposed Julia terminal editor is developed considering the modular interaction. The developed package operates in multiple modes such as edit, file management, execution, debug, and syntax mode, where user can freely hover in required mode of interest making it a context-aware package. The different modes of operation are provided in key bindings that the user can use and are readily available in specified mode with the information of inputs which reduces cognitive friction. The following section discusses the code organisation, applicability, and the working of the proposed terminal editor in different modes. In addition, the comparison of the proposed package with the existing IDE and code editor is carried out.

### A. Code organization

The proposed package is natively built on Julia programming language version 1.11.5 with the windows system of 4GB RAM, intel 7, 9th generation, 64-bit operating system, x64-based processor with clock frequency of 1.88 GHz – 2 GHz. MinTEJ is independent of third-party packages that reduces the software over-head caused due to external packages. The codes are organized in the source folder and consists of the separate Julia codes for each mode and every option. The main code is organized in the Julia module structure and exports the function of terminal editor. Functionalities of terminal editor are organized to hover under different options based on the user input. The detailed package organisation for different modes with a corresponding list of files along with their functionality is provided in Appendix-A. The presented code organization provides modularity and flexibility to organise and add future modifications for code reusability and extensibility. The following section provides the discussion on various modes of operations present in MinTEJ.

```
julia> using Main.commands_editor

julia> mintej()

++++++++++++++++++++++++++++++++++++++++++++++++++++++++
Welcome to Minimalistic Julia Terminal Editor Version 00
++++++++++++++++++++++++++++++++++++++++++++++++++++++++
Enter in file management system >>fms
Enter in Editor mode >>edm
Enter in Execution mode >>exe
Enter in debug mode >>db
Enter in syntax mode >>syntx
MinTEJ>>
```

Figure 3. MinTEJ Mode: Activated by function mintej().

```
MinTEJ>>edm

Enter the file name

edm>>info
w            :: Writes the file
rd_lines     :: Reads the file
d            :: deletes the lines in file
cp           :: Copies the lines in file
bs           :: Adds blanck lines in file
cm           :: Adds comments to lines in file
uncm         :: Uncomments the lines in file
onfile       :: User enter the filename
clear        :: Clears the console
edm>>
```

Figure 4. Options in editor mode.

```
edm>>onfile

Enter the file name
myfirstcode.jl
Choose option from info list
edm>>w
Editing: myfirstcode.jl
1:
Type new lines. Type 's' to save and exit.
global x = 0
while x <= 5
      global x = x + 1
      println("The number is:",x)
end
s
Saved to myfirstcode.jl
edm>>
```

Figure 5. Option onfile to create a new file or to open existing file.

```
edm>>rd_lines

Reading: myfirstcode.jl
1:
2:global x = 0
3:while x <= 5
4:      global x = x + 1
5:      println("The number is:",x)
6:end

Info: Enter Start_line, off-set before the start line, End_line till to be read
1,0,3

1:
2:global x = 0
3:while x <= 5
edm>>
```

Figure 6. Options to read the file on console. The file can be read selectively using the parameters start line, offset, and till a specific line number.

### *B. Terminal modes in proposed Julia package - MinTEJ*

The proposed terminal editor for Julia is designed to have minimalistic approach followed by its modular and flexible design paradigm. In MinTEJ, the user can switch in different operation mode with interactive engagement with the provided functionalities. The following section discusses the operations and functionality present in various modes of operation.

#### *1. Editor mode*

The program is initiated using the module Main.Commands_editor, and Julia terminal editor function mintej() is exported in Julia REPL environment. The program is triggered in the MinTEJ editor which serves access to different modes by entering the keyword, such as *fms* for file management mode, *edm* for editor, *exe* for execution, whereas debugger and syntax mode can be entered using keywords such as *db* and *syntx* as shown in Figure 3. Every mode has the option "info" that provides the information about the modes and its relevant options, helping user to access functionalities with less cognitive effort. Info option for edm is shown in Figure 4 which provides quick access to functionality in edm mode. In editor mode, the user is prompted to enter the file name that needs to be edited or the file that needs to be created, as seen in Figure 4. Files are created with the dot jl extension i.e., filename.jl. To demonstrate the editor capability, a file with name myfirstcode.jl is created. In case, a user wants to modify the existing file then use of option "onfile" where user can enter the file name of interest as in Figure 5. This provides the flexibility to user to edit and load the desired files. The developed editor provides an interactive mode to edit files that the user is backed by multiple options. Using option *onfile* prompt, the user accesses a file name, if does not exist in a system, the program creates a new file and if exist then file is ready for the operation. The user can edit the file using the option "*w*" which stands for write as shown in Figure 5. In Figure 5, a code is written which has a simple loop structure, and it is saved in the file with option *s*, further displaying a message saved to myfirstcode.jl. The files can be read back on the console using the option *rd_lines* and this option has a selective ability to read a line by setting the parameters such as start line number and offset along with end line number. This provides a feature to read a code in selective way. The example of usage of readline is provided in Figure 6.

Deleting the content of files is one of the essential features of a terminal editor. The editor has a delete mode d to delete the specific lines in the code. This option has selective features in which the user can select the set of lines or individual lines. The program reads the file on console and prompts the information (or instruction) to the user to enter the line numbers to be deleted. The user can provide input such as 4:7 which indicates the line number 4, 5, 6 and 7 should be deleted. In Figure 7, the application of *d* option is demonstrated, the user set line number as 1 and the program executes to delete the line number 1. The copying option in file editing mode is provided by option *cp*. The file is re-read on screen and the info to copying the line number is prompted on the display. Like *d* option *cp* prefers the input to individual or multi-line, followed by the locations of the line to copy. The example of copying the content is demonstrated in Figure 8. Figure 9 and Figure 10 demonstrate adding a blank line and this option is enabled by *bs*. Similarly, in *bs* command the program prompts the user to enter the line number for adding a blank line. The user can provide the input in the form of single number or multiple line numbers, same as in option *d*. Figure 10 shows that the blank line is inserted at line no. 4 which is input provided by user as shown in Figure 4.

The comment and uncomment functionality are demonstrated in Figure 11 - Figure 14. In editor mode, commenting and uncommenting are provided using options *cm* and *uncm*. Like other options in *edm*, the file is read, and user can comment and uncomment using the line number or multiple lines. The *edm* is interactive and guided by the prompts so user seamlessly edit the Julia file. This terminal approach presented in this work is modular and purely relies on terminal operations. MinTEJ offers file management such as file & directory navigation, creating & deleting files & folder, tree like viewing and comparing the files, and is discussed in the below section.

```
edm>>d

Re-reading saved file:
1:
2: global x = 0
3: while x <= 5
4:       global x = x + 1
5:       println("The number is:",x)
6: end

Info: Enter Start_line & End_line number or line to be deleted
1
1: global x = 0
2: while x <= 5
3:       global x = x + 1
4:       println("The number is:",x)
5: end
Saved to myfirstcode.jl
```

Figure 7. Demonstration of delete option d. (MinTEJ prompt to enter the specific line number. Here, user enters line number as 1. The first line 1: is deleted.)

```
edm>>cp

Re-reading saved file:
1: global x = 0
2: while x <= 5
3:       global x = x + 1
4:       println("The number is:",x)
5: end

Info: Enter line number to be copied
2:5

Info: Enter location to be copied
6
1: global x = 0
2: while x <= 5
3:       global x = x + 1
4:       println("The number is:",x)
5: end
6: while x <= 5
7:       global x = x + 1
8:       println("The number is:",x)
9: end
Saved to myfirstcode.jl
edm>>
```

Figure 8. Execution of copying options. (The user is prompted to enter the line number to copy and the location where to paste it.)

```
edm>>bs

Re-reading saved file:
1: global x = 0
2: while x <= 5
3:       global x = x + 1
4:       println("The number is:",x)
5: end
6: while x <= 5
7:       global x = x + 1
8:       println("The number is:",x)
9: end

Info: Enter line number where the blank space is needed
4
```

Figure 9. Demonstration of blank line insertion. (The program provides the prompt to enter line number. Here, users select line number 4.)

```
1: global x = 0
2: while x <= 5
3:       global x = x + 1
4:
5:       println("The number is:",x)
6: end
7: while x <= 5
8:       global x = x + 1
9:       println("The number is:",x)
10: end
Saved to myfirstcode.jl
```

Figure 10. Blank line is inserted at line 4 inline with user input.

```
edm>>cm

Re-reading saved file:
1: global x = 0
2: while x <= 5
3:       global x = x + 1
4:
5:       println("The number is:",x)
6: end
7: while x <= 5
8:       global x = x + 1
9:       println("The number is:",x)
10: end

Info: Enter line number to comment
7:10
```

Figure 11. Execution of comment option. (The users file can be commented using the option cm and program prompt the line number to comment as the input.)

```
7:10
1: global x = 0
2: while x <= 5
3:       global x = x + 1
4:
5:       println("The number is:",x)
6: end
7: #while x <= 5
8: #      global x = x + 1
9: #      println("The number is:",x)
10: #end
Saved to myfirstcode.jl
edm>>
```

Figure 12. Demonstration of multi-line comment option. The user has provided line number as 7:10, which indicates the user wants to comment line number 7, 8, 9, 10.

```
edm>>uncm

Re-reading saved file:
1: global x = 0
2: while x <= 5
3:       global x = x + 1
4:
5:       println("The number is:",x)
6: end
7: #while x <= 5
8: #      global x = x + 1
9: #      println("The number is:",x)
10: #end

Info: Enter line number to uncomment
7:10
```

Figure 13. Execution of the uncommented option with the given line numbers. (Similar to comment command, the editor mode has option uncm for uncommenting the line. With the input option, program prompts the line number to be uncommented.)

```
1: global x = 0
2: while x <= 5
3:       global x = x + 1
4:
5:       println("The number is:",x)
6: end
7: while x <= 5
8:       global x = x + 1
9:       println("The number is:",x)
10: end
Saved to myfirstcode.jl
edm>>
```

Figure 14. Execution of multi-line uncomment. (The user provide option to uncomments are line number 7, 8, 9, 10 and the program removes the additional character # which is used in Julia programming language to uncomment and comment the program.)

*2. File management system mode*

File management system in the proposed MinTEJ uses Julia specific semantics for handling various file management operations such as listing the file, changing directory and many more operations. The *fms* approach facilitates the user to manages their files and directory in Julia terminal. User need to use the keyword *fms* to manage the file operations. The various options present in *fms* can be listed using the keyword as "info" as shown in Figure 15. *fms* mode has flexibility to use the options inter-changeably that allows users to manage the files effectively. The user can access the current working directory by using the command *lc* and conveniently change the working directory using command *cdir* as shown in Figure 16. In Figure 17, with input option as *cdir*, the program prompts the user to enter the path for changing directory and current directory is printed over the console. The option *ls* is used to list out the content of files and directories present in the user's path. The *fms* is a feedback-based system that makes user more convenient for working with various Julia files. During copying option *cpy* the program prompts the source name to be entered and destination name for copying. The presented *cpy* option also takes the full-path, making it more convenient to work around different directories and files. Figure 18 shows the sample file for copying from folder test4 to test3. The *fms* mode has options to create a new folder using *mkdr* (make directory). With the input option as mkdr as seen in Figure 19, the program emits feedback to add the folder name and user can add the path or can create folder locally in working directory. The file/directory creations are important aspect for user to manage and organize the files.

In the proposed package, *fms* mode has option to delete file and option to delete a folder as *delfl* and *deldir*. As the major of safety, during deleting the file/folder program issues a warning and prompts for folder path. In case, user does not want to proceed with the *delfl* or *deldir* then it can be skipped by hitting RETURN/ENTER button. The demonstration of *deldir* is shown in Figure 20, the directory test4 is deleted and the confirmation can be done by listing out with command *ls*. The *fms* mode has option to rename the file and folder using *rn*. When program detects the key *rn*, the program prompts a message to enter the old file/folder name, followed by new file/folder name. The usage of option *rn* is shown in Figure 21 and Figure 22 to rename the file test_file3.jl and test_file4.jl and

```
MinTEJ>>fms
fms>>info
lc     :: Current directory
dir    :: List directory
ls     :: List directory with path
cdir   :: Change directory
cpy    :: Copy files/directory
delfl  :: delete directory
deldir:: delete directory
mkdr   :: Create directory
rn     :: Rename file/ directory
clear :: clear console
tree   :: View tree structure
rd_lines :: Reads file
cmp :: Compares two text files
cmdwin :: Opens the windows instance
       :: Unknown command
fms>>
```

Figure 15. Options in fms mode

```
fms>>lc

Printing the current directory
C:\Users\Tejasvi\Desktop\Julia_Program\MinTEJ
fms>>cdir

Changing directory

Enter the path
C:\Users\test
C:\Users\test
fms>>dir

Printing the list content of directory
test1
test2
test3
test_file1.jl
test_file2.jl
test_file3.jl
```

Figure 16. Demonstration of listing and change of directory in fms mode.

```
fms>>cdir

Changing directory

Enter the path
C:\Users\Tejasvi\Desktop\Julia_Program\MinTEJ\test
C:\Users\Tejasvi\Desktop\Julia_Program\MinTEJ\test
fms>>ls

List all files and directory with path
C:\Users\Tejasvi\Desktop\Julia_Program\MinTEJ\test\test1
C:\Users\Tejasvi\Desktop\Julia_Program\MinTEJ\test\test2
C:\Users\Tejasvi\Desktop\Julia_Program\MinTEJ\test\test4
C:\Users\Tejasvi\Desktop\Julia_Program\MinTEJ\test\test_cmp_1.jl
C:\Users\Tejasvi\Desktop\Julia_Program\MinTEJ\test\test_cmp_2.jl
C:\Users\Tejasvi\Desktop\Julia_Program\MinTEJ\test\test_file1.jl
C:\Users\Tejasvi\Desktop\Julia_Program\MinTEJ\test\test_file2.jl
C:\Users\Tejasvi\Desktop\Julia_Program\MinTEJ\test\test_file3.jl
fms>>
```

Figure 17. Usage of change directory using cdir

```
fms>>cpy

Enter the source file/folder name
test4

ENter the destination file/folder name
test3
Copied 'test4' to 'test3'
```

Figure 18. Copying the content of file and folder to defined destination.

```
fms>>mkdr

Add the folder name to create directory
C:\Users\Tejasvi\Desktop\Julia_Program\MinTEJ\test\test4
fms>>ls

List all files and directory with path
C:\Users\Tejasvi\Desktop\Julia_Program\MinTEJ\test\test1
C:\Users\Tejasvi\Desktop\Julia_Program\MinTEJ\test\test2
C:\Users\Tejasvi\Desktop\Julia_Program\MinTEJ\test\test3
C:\Users\Tejasvi\Desktop\Julia_Program\MinTEJ\test\test4
C:\Users\Tejasvi\Desktop\Julia_Program\MinTEJ\test\test_file1.jl
C:\Users\Tejasvi\Desktop\Julia_Program\MinTEJ\test\test_file2.jl
C:\Users\Tejasvi\Desktop\Julia_Program\MinTEJ\test\test_file3.jl
fms>>
```

Figure 19. Make directory is demonstrated by option mkdr

```
fms>>deldir

Warning: Provide the folder path
C:\Users\Tejasvi\Desktop\Julia_Program\MinTEJ\test\test4

Folder is deleted!
fms>>ls

List all files and directory with path
C:\Users\Tejasvi\Desktop\Julia_Program\MinTEJ\test\test1
C:\Users\Tejasvi\Desktop\Julia_Program\MinTEJ\test\test2
C:\Users\Tejasvi\Desktop\Julia_Program\MinTEJ\test\test3
C:\Users\Tejasvi\Desktop\Julia_Program\MinTEJ\test\test_file1.jl
C:\Users\Tejasvi\Desktop\Julia_Program\MinTEJ\test\test_file2.jl
C:\Users\Tejasvi\Desktop\Julia_Program\MinTEJ\test\test_file3.jl
fms>>
```

Figure 20. Demonstration of delfl and deldir. (fms mode has the deleting option for file and directory. These options are activated keys.)

```
fms>>rn

Enter the oldname file/folder name
test_file3.jl

Enter the newname file/folder name
test_file4.jl
Renamed 'test_file3.jl' to 'test_file4.jl'
fms>>ls

List all files and directory with path
C:\Users\Tejasvi\Desktop\Julia_Program\MinTEJ\test\test1
C:\Users\Tejasvi\Desktop\Julia_Program\MinTEJ\test\test2
C:\Users\Tejasvi\Desktop\Julia_Program\MinTEJ\test\test3
C:\Users\Tejasvi\Desktop\Julia_Program\MinTEJ\test\test_file1.jl
C:\Users\Tejasvi\Desktop\Julia_Program\MinTEJ\test\test_file2.jl
C:\Users\Tejasvi\Desktop\Julia_Program\MinTEJ\test\test_file4.jl
fms>>
```

Figure 21. fms mode to rename the file.

```
fms>>ls

List all files and directory with path
C:\Users\Tejasvi\Desktop\Julia_Program\MinTEJ\test\test1
C:\Users\Tejasvi\Desktop\Julia_Program\MinTEJ\test\test2
C:\Users\Tejasvi\Desktop\Julia_Program\MinTEJ\test\test3
C:\Users\Tejasvi\Desktop\Julia_Program\MinTEJ\test\test_file1.jl
C:\Users\Tejasvi\Desktop\Julia_Program\MinTEJ\test\test_file2.jl
C:\Users\Tejasvi\Desktop\Julia_Program\MinTEJ\test\test_file3.jl
fms>>rn

Enter the oldname file/folder name
test3

Enter the newname file/folder name
test4
Renamed 'test3' to 'test4'
fms>>
```

Figure 22. Rename option using rn for directory.

```
fms>>tree

 List tree
□  C:\Users\Tejasvi\Desktop\Julia_Program\MinTEJ\test
  └── test_file1.jl
  └── test_file2.jl
  └── test_file3.jl
□  C:\Users\Tejasvi\Desktop\Julia_Program\MinTEJ\test\test1
  └── syntax_fcn.jl
□  C:\Users\Tejasvi\Desktop\Julia_Program\MinTEJ\test\test2
  └── syntax_fcn_1.jl
□  C:\Users\Tejasvi\Desktop\Julia_Program\MinTEJ\test\test4
  └── syntax_fcn_3.jl
fms>>
```

Figure 23. Tree option to visualize the content of directory and sub-directory.

```
fms>>cmp

Enter the text file 1
test_file1.jl

Enter the text file 2
test_file2.jl
Files are identical
File name:         test_file1.jl                 | test_file2.jl
File size (KB):    10.26                         | 10.26
Total lines:       383                           | 383
Total words:       383                           | 383
fms>>
```

Figure 24. Comparing file in fms mode with identical files.

directory as test3 to test4. The *fms* also has *tree* option which list out the content of directory and its sub-directories. The usage of tree option is shown in Figure 23. Maintaining the files during the development mode is crucial, and users may often require the code files that need to be compared. Therefore, the developed *fms* mode is equipped with the option to compare the files, and this file comparison is based on character and word comparison. The use case of comparing the files using the option *cmp* is provided in Figure 24 and Figure 25. Program prompt the name of file that needs to be compared and if the files are identical, the fms provide feedback that the files are identical else, it flashes a message saying "files do not match". The additional information for file comparison includes the file size, number of lines and total number of words. The user can find the mismatch by accessing logs printed on the screen and the changes in the additional information such file size, total number of lines and words. The *fms* mode have functionality of read lines *rd_lines* to read the input file which operates the same as *edm* mode. It has a capability to trigger multiple working windows with command *cmdwin*. Currently, *cmdwin* is tested for windows machines.

### 3. *Execution mode*

The developed MinTEJ package provides the code execution in the working environment. As mentioned, the available option in *exe* mode can be viewed by using keyword *info* and provides flexibility of traversing from one option to another. The *exe* mode is directed through MinTEJ and various options such as execution *e*, and REPL like session *repl* can be initiated in this mode. The demonstration of these options and execution of program file is shown in Figure 27. While using this, it prompted to enter the file name for execution after entering the input option. Once the file name is entered then the program

```
fms>>cmp

Enter the text file 1
test_cmp_1.jl

Enter the text file 2
test_cmp_2.jl
File_A: 2 println("Hello world")
File_B: 2 println("hello Julia Programming")
File_A: 2 println("Hello world")
File_B: 2 println("hello Julia Programming")
Files do not match
File name:         test_cmp_1.jl                 | test_cmp_2.jl
File size (KB):    0.02                          | 0.04
Total lines:       2                             | 2
Total words:       2                             | 2
fms>>
```

Figure 25. Comparison of non-identical files.

executes and the output of program involving the print statements are shown on the console. The sample program "myfirstcode.jl" shown in Figure 26 is used to demonstrate the code execution. The standard output and standard error are directed into text file after the code execution, and error logs are saved into a log file generated for reference. These log files are available in working directory. To showcase the error capturing feature of *exe* mode, a sample code is considered as given in Figure 28, where the code has a missing double code in print command. This code is executed with exception handling to capture the error and the error is shown on the console as shown in Figure 29. To activate the REPL like execution session, keyword *repl* is used as in shown in Figure 30 and thereafter, the interactive code execution is performed for quick understanding of the typed code behaviour. With *repl* as an input to *exe* mode, the user is prompted with the option to either read Julia file or to work in repl like session. To exit from the *repl* session the user should type desired code followed by END". Once "END" is detected, then execution is performed" and redirected to the output. This *repl* execution option mode allow users to quickly assess the code or logic.

```
MinTEJ>>fms
fms>>rd_line
Unknown command
fms>>rd_lines

Enter the file name
myfirstcode.jl

Reading: myfirstcode.jl
1:global x = 0
2:while x <= 5
3:      global x = x + 1
4:      println("The number is:",x)
5:end
6:
```

Figure 26. Programme file for code execution.

```
MinTEJ>>exe
exe>>info
e              :: Execution of julia file
repl           :: REPL like session
info           :: command information
exe>>e
exe>>
Enter the file name to execute
myfirstcode.jl


The number is:1
The number is:2
The number is:3
The number is:4
The number is:5
The number is:6

exe>>
```

Figure 27.Options in exe mode with execution results of program (myfirstcode.jl).

```
fms>>rd_lines

Enter the file name
test_cmp_2.jl

Reading: test_cmp_2.jl
1:
2:println("hello Julia Programming)
```

Figure 28. Demonstration of rd_line option. (File tmp_cmp_2.jl with incorrect syntax – missed double codes.)

```
MinTEJ>>exe
exe>>e
exe>>
Enter the file name to execute
test_cmp_2.jl

whoops - No file or Program runs an issue to execute
caught exception:LoadError("C:\\Users\\Tejasvi\\Desktop\\Julia_Program\\
MinTEJ\\test\\test_cmp_2.jl", 2, Base.Meta.ParseError("ParseError:\n# Er
ror @ C:\\Users\\Tejasvi\\Desktop\\Julia_Program\\MinTEJ\\test\\test_cmp
_2.jl:2:35\n\nprintln(\"hello Julia Programming)\n#
          \n└ ── unterminated string literal", Base.JuliaSyntax.ParseEr
ror(Base.JuliaSyntax.SourceFile("\nprintln(\"hello Julia Programming)\n"
, 0, "C:\\Users\\Tejasvi\\Desktop\\Julia_Program\\MinTEJ\\test\\test_cmp
_2.jl", 1, [1, 2, 36]), Base.JuliaSyntax.Diagnostic[Base.JuliaSyntax.Dia
gnostic(36, 35, :error, "unterminated string literal"), Base.JuliaSyntax
.Diagnostic(36, 35, :error, "Expected `)`")], :string)))
exe>>
```

Figure 29. Error handling using exception handling approach.

### 4. *Debug mode*

The debug mode has multiple approaches to debug the code in an interactive way i.e., using breakpoint approach or line-by-line approach (i.e., REPL session). The available options in debugging mode are shown in Figure 31. Use of breakpoint and stepin is shown in Figure 31. The suffix "*i*" in *ibp* and *istepin* indicates the interactive operations. The debug mode provides the REPL like session for checking the code and for quick edit over terminal. The *db* mode has option as *bp* which stands for break point and *stepin* mode to step into code in controlled way. The interactive mode in bp is activated using ibp. The sample code is written on terminal as shown in Figure 32 in REPL like session and code is reproduced on screen for user which can be used as verification. The program prompts the location of break point to be added as shown in Figure 33 and the code pause at break point say line number 5 and the environment variables are read for the user's introspection of the code. The program in debugging mode is continued until RETURN keyword is entered as in Figure 34. The functionality like stepin using AST traversal is wrapped into macros and unlike the breakpoint functionality, the stepin works in interactive as well as file mode. In file mode, user can provide the Julia code for introspection. Figure 35 demonstrates the usage of step functionality using the keyword *istepin* for REPL like session. The sample code for using 'for' loop to increment the value over iterations is considered to understand the stepin functionality. Once the code is entered by user, the program emits the evaluation of expressions that are traced using AST. The program provides the provision to inspect the environment variables. In each iteration/decision loop, the environment variable values are traced, and option for entering the new variables are provided for introspection of environment

```
exe>>repl

Type jl to read file

OR

Hit return to work in REPL like session

Enter your code. Type 'END' on a new line to finish:
exe>>for k = 1:3
exe>>println(k, ":hello Julia Programming")
exe>>end
exe>>END
1:hello Julia Programming
2:hello Julia Programming
3:hello Julia Programming
Result: nothing
exe>>
```

Figure 30. REPL like interaction

variables. The solution of istepin is presented in Figure 36 which tracks the global variables $x$ and $x + 1$ as code flows. User can use existing code present in dot jl file to read into stepping in function as shown in Figure 37. The program prompts the filename that can be used as input code to step into function. In istepin mode, the program emits the expression as program evolves to its completion and the option to monitor environment variable over code evaluation is prompted over the terminal. The proposed debug mode is minimalistic and allows quick inspection of the codes providing user flexibility to use different mode either using file or REPL likes session.

### 5. *Syntax mode*

The syntax mode is a knowledge database in which the syntax and examples of commonly used commands are included. In addition, users can store the references for the code and this mode is activated through MinTEJ using *syntx* keyword. The list of available databases is traced using command *info* as shown in Figure 38. The syntx mode is evoked

```
MinTEJ>>db
db>>info
db          :: debug mode - step and break
bp          :: Break point mode
ibp         :: interactive break
            :: file mode break
stepin      :: Run the code step wise
            :: Run file code step wise
istepin     :: interactive step
clear       :: clears the console
db>>
db>>
db>>
```

Figure 31. Options in debug mode.

```
db>>bp
Type ibb for interactive breakpoint
Hit RETURN for adding breakpoint in julia file
ibb
db>>function mi()
db>>x = 0
db>>for k = 1:3
db>>x = x+1
db>>end
db>>end
db>>mi()
db>>END
1:  function mi()
2:  x = 0
3:  for k = 1:3
4:  x = x+1
5:  end
6:  end
7:  mi()
```

Figure 32. ibp: Interactive breakpoint in REPL session.

```
Add the break point
5
Breakpoint hit at line
Variables in scope:
  k = 1
  x = 1

Breakpoint hit at line
Variables in scope:
  k = 2
  x = 2

Breakpoint hit at line
Variables in scope:
  k = 3
  x = 3

db>>
```

Figure 33. Stepwise run when a break point is hit.

```
db>>bp
Type ibb for interactive breakpoint
Hit RETURN for adding breakpoint in julia file

Enter the file name
test_my.jl
Add the break point
6
Breakpoint hit at line
Variables in scope:
  k = 1
  x = 1

Breakpoint hit at line
Variables in scope:
  k = 2
  x = 2

Breakpoint hit at line
Variables in scope:
  k = 3
  x = 3

db>>
```

Figure 34. Adding breakpoint in Julia file.

```
istepin
db>>global x = 0
db>>for k = 1:2
db>>global x
db>>x = x + 1
db>>end
db>>END
1:  global x = 0
2:  for k = 1:2
3:  global x
4:  x = x + 1
5:  end
```

Figure 35. Interactive stepping function in REPL like session.

```
Line at $(Expr(:escape, 2)): global x = 0
Enter variable assignment (e.g., list_variables = [:x, :y]): list_variables = [:x]
Entering for loop at depth $(Expr(:escape, 2)): for k = 1:2
    #= C:\Users\Tejasvi\Desktop\Julia_Program\MinTEJ\output_debug.jl:4 =#
    global x
    #= C:\Users\Tejasvi\Desktop\Julia_Program\MinTEJ\output_debug.jl:5 =#
    x = x + 1
    #= C:\Users\Tejasvi\Desktop\Julia_Program\MinTEJ\output_debug.jl:6 =#
end
Line at $(Expr(:escape, 4)): global x
  x = 0
Line at $(Expr(:escape, 4)): x = x + 1
  x = 0
Enter variable assignment (e.g., list_variables = [:x, :y]):
  x = 1

Line at $(Expr(:escape, 4)): global x
  x = 1
Line at $(Expr(:escape, 4)): x = x + 1
  x = 1
Enter variable assignment (e.g., list_variables = [:x, :y]):
  x = 2
```

Figure 36. Output function of the stepping function.

```
db>>stepin
Type istepin for interactive stepping
Hit RETURN for adding iteractive stepping in julia file

Enter the file name
test_istep.jl
Line at $(Expr(:escape, 2)): global x = 0
  x = 2
Entering for loop at depth $(Expr(:escape, 2)): for k = 1:2
    #= C:\Users\Tejasvi\Desktop\Julia_Program\MinTEJ\test\output_debug.jl:5 =#
    global x
    #= C:\Users\Tejasvi\Desktop\Julia_Program\MinTEJ\test\output_debug.jl:6 =#
    x = x + 1
    #= C:\Users\Tejasvi\Desktop\Julia_Program\MinTEJ\test\output_debug.jl:7 =#
end
Line at $(Expr(:escape, 4)): global x
  x = 0
Line at $(Expr(:escape, 4)): x = x + 1
  x = 0
Enter variable assignment (e.g., list_variables = [:x, :y]): list_variables = [:x]
  x = 1

Line at $(Expr(:escape, 4)): global x
  x = 1
Line at $(Expr(:escape, 4)): x = x + 1
  x = 1
Enter variable assignment (e.g., list_variables = [:x, :y]):
  x = 2
```

Figure 37. File mode stepping into program with controlled execution

```
MinTEJ>>syntx
syntx>>info
info
---------------------
Examples are taken from
1. https://juliabyexample.helpmanual.io/
2. https://www.datacamp.com/cheat-sheet/julia-basics-cheat-sheet

:print
:arithmatic
:parse
:formatting
:stringformating
:errorhandling
:arrays
:dictionary
:variables
:for
:if
:while
:function
:mathematicalprogramming
```

Figure 38. Syntax mode option.

by keywords such as print, arithmetic, while and many more. Keywords in syntax modes provides the knowledge base and a reference for programmers. The user has a provision to update syntax mode file to add new references or knowledge base. Figure 39 demonstrates the example of mathematical programming and syntax mode is evoked by keyword mathematical-programming in this example.

### C. *Comparison and performance analysis*

The memory consumption of application is important aspect to understand the memory-overhead of the program. To understand the memory usage of the proposed editor, a comparative study considering MinTEJ and other popular editors such as VS code and Notepad++ is studied. This simulation study comprises of memory consumption in idle condition and during program execution. The memory profile is calculated using the moving average over the buffer size of 4. The memory consumption during program execution is calculated using psutil package based python program. At every 5 seconds, the stastistics from the system are collected and these data are filtered using the moving average method to remove the noise in the memory measurement. The idle condition memory consumption profile of editors such as Notepad++, VS code, and proposed MinTEJ is shown in Figure 40. It is also worth to note the Figure 40 shows the memory profile of Julia. The MinTEJ is completely written in Julia, at the idle condition, the memory consumption of the program as shown in Figure 40 consumes 120 MB. The Julia program in terminal shows the 210 MB memory usage at the beginning and then slowly drops down when the Julia goes in idle state for long time. The VS code is an advanced feature IED which has higher memory consumption at the beginning of the application because of multiple dependencies such as language server, plugin and many more. The VS code requires almost 800 MB at the start and with time it settles to 400 MB. Similarly, the notepad++ memory burden is around 40 to 50 MB at start and may go low after some time. It shows that Notepad++ requires less memory overhead and is lightweight editor in idle conditions. The proposed application is Julia based, which is lightweight, minimalistic terminal editor compared to VS code, and can be seen on slight higher than Notepad++. As matter of fact, notepad++ and VS code are much developed editor, however the proposed package offers a simplistic and robust approach for Julia Programmers. Figure 41 shows the memory consumption during program execution using the proposed work package in comparison to Notepad++, VS code, and Julia. The test program consists of loading memory buffer in Julia.

Every buffer is loaded with averagely 8 MB and simulation iterations are set to 500. During the program execution, the memory consumption of Julia increases filling up these buffer. The test code with the description is provided in the Appendix-B. From Figure 41, it observed that the code is executed in Julia terminal as well as MinTEJ; The memory profile shows the memory consumption in Julia and MinTEJ increases to peak point and towards program end, it drops. Similar observation made while running program on VS code and Notepad++. The VS code has higher memory burden and uses around 500 MB. Here, the VS code sets the Julia execution as separate process which marked in Figure 41 as VS Code + Julia. The sum of total memory set VS code and Julia process at peak is around 4000 MB. Similarly, it is observed that notepad++ runs the Julia

```
syntx>>mathematicalprogramming
Mathematical Optimization in Syntax
More examples can be found here:
https://jump.dev/JuMP.jl/stable/tutorials/getting_started/getting_started_with_JuMP/
-----------------------------------------
using JuMP
using HiGHS
model = Model(HiGHS.Optimizer)

@variable(model, x >= 0)
@variable(model, 0 <= y <= 3)
@objective(model, Min, 12x + 20y)

@constraint(model, c1, 6x + 8y >= 100)
@constraint(model, c2, 7x + 12y >= 120)

print(model)
optimize!(model)

termination_status(model) #OPTIMAL::TerminationStatusCode = 1

primal_status(model) #FEASIBLE_POINT::ResultStatusCode = 1

dual_status(model) #FEASIBLE_POINT::ResultStatusCode = 1

objective_value(model)
value(x)
value(y)
shadow_price(c1)
shadow_price(c2)
```

Figure 39. Example of syntax mode with mathematical programming.

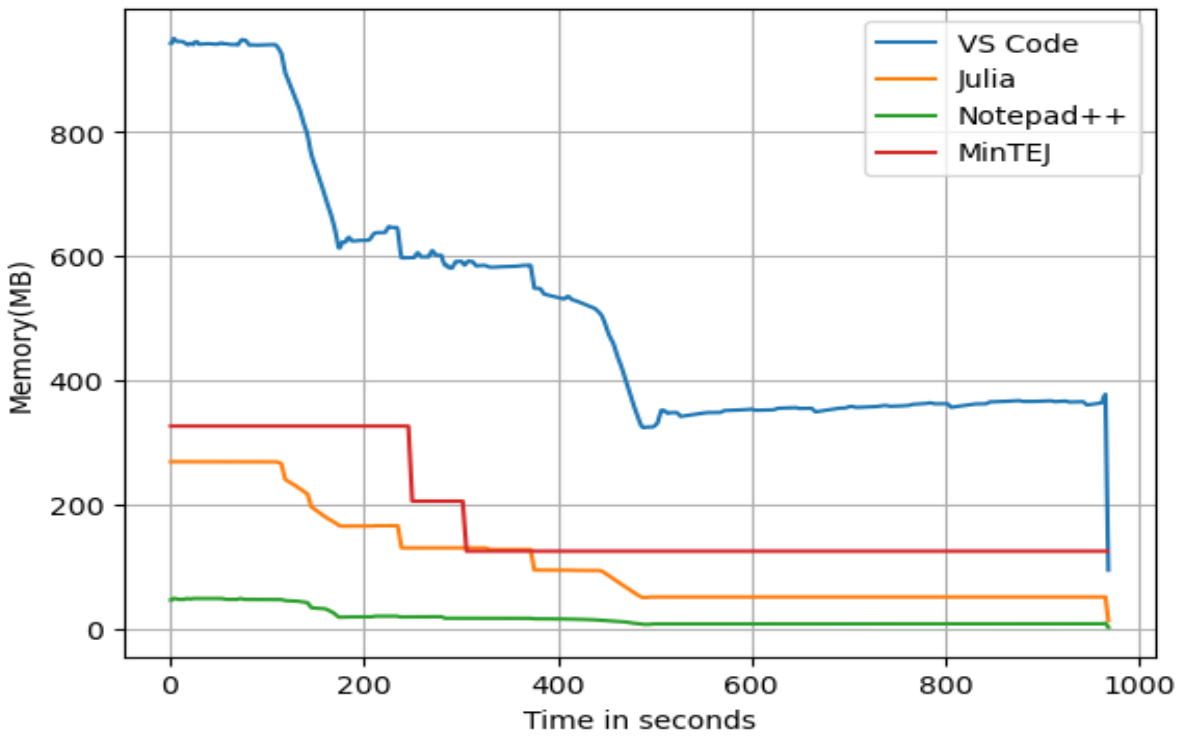

Figure 40. Memory consumption data for the VScode, Julia, Notepad++, MinTEJ when the applications are idle. (MinTEJ is Julia package, the idle memory is 120 MB.)

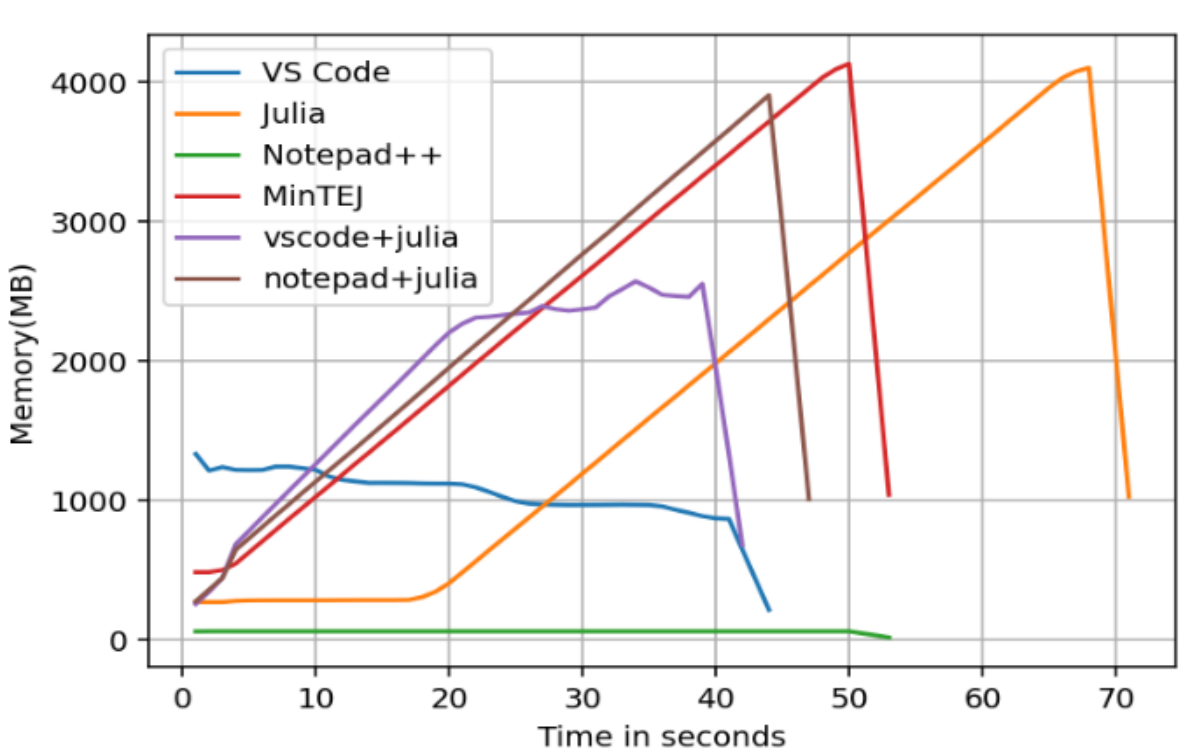

Figure 41. Memory consumption of the Vs code, Notepad++, Julia, MinTEJ while program execution. (VS code+Julia indicates the process program execution using Vs code initiated in Julia, similarly notepad+Julia is the process initiated by notepad++ when the program is running.)

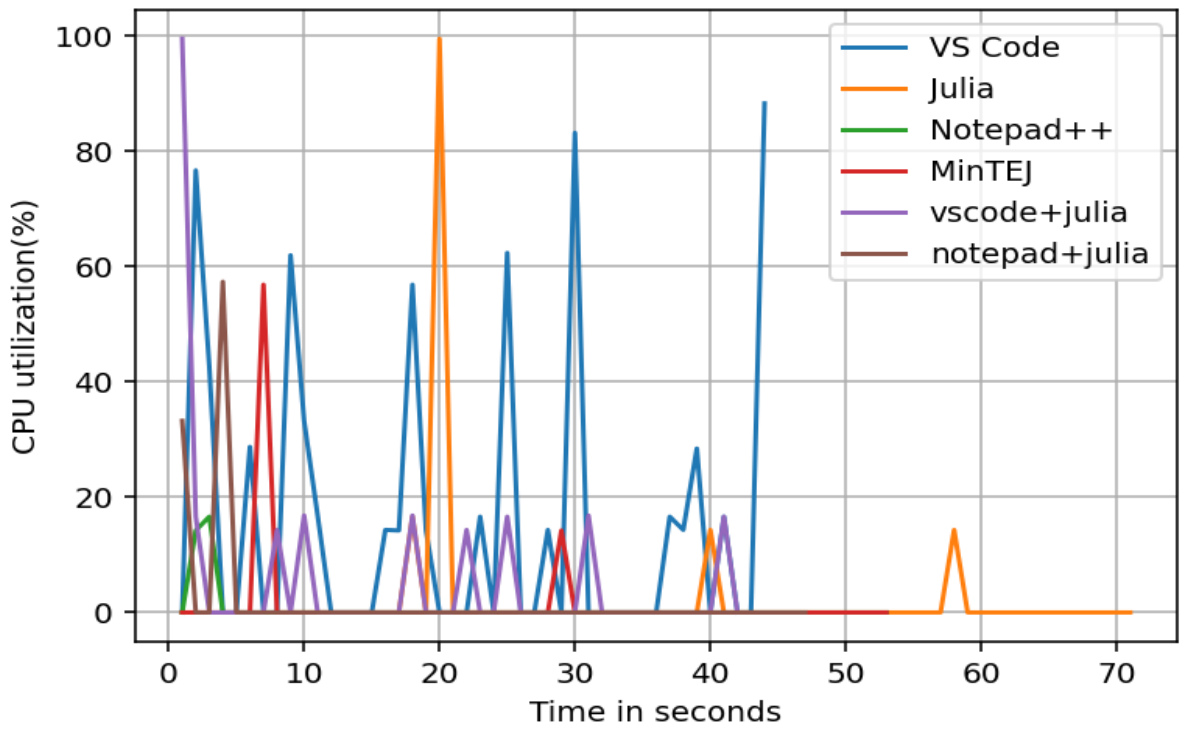

Figure 42. CPU utilization percentage in VS code, Notepad++ and the proposed package MinTEJ

using the npp execution with external binding. The memory consumption of Notepad++ is approximately 60 MB. The proposed MinTEJ has less memory burden at start as compared to VS Code. Notepad++ is pure text editor but with external binding the code is executed in Julia terminal. The proposed editor have memory overhead compared to Notepad++. However, the provided comparison is not direct but provides the fair memory utilization idea of exisitng editors and the proposed package. The CPU percentage utilization provides information of resource utilization. The percentage CPU utilization of VS code, Notepad++ and the developed MinTEJ along with Julia execution as process is shown in Figure 42. It is observed that CPU utilization of VS code with the Julia process is comparatively higher compared to Notepad++ and MinTEJ. It is also observed that Notepad++ with Julia process shows a spike in CPU usage when Julia process is executed. The MinTEJ requires less resource overhead compared to mentioned editors that makes MinTEJ a light weight minimalsitic editor.

## IV. Conclusion

This work proposes the sequential modal interaction architecture (SMIA) based minimalistic terminal editor for Julia i.e., MinTEJ operating in different modes such as editor mode, file management mode, execution mode, debug mode, and syntax mode. It is designed based on modular approach and provides the flexibility to Julia users. These approaches are simple and well-defined providing the users a control upon the code and its maintenance. The proposed integrated editor MinTEJ natively uses the Julia programming language for developing the minimalistic editor, making it robust and less resource dependent. The proposed packages do not use any third-party Julia packages to reduce the dependencies of resource overhead. MinTEJ is a guided mode editor with prompts and commands that are readily available for users. The file management provides the flexibility to maintain the code and navigate to different files and folders of the computer. The proposed package also serves execution and debugger mode using the abstract syntax tree approach. The execution mode also provides the REPL like session to execute the code for quick check. The proposed MinTEJ package is tested for its memory usage and CPU utilization with existing editors. The results show that the MinTEJ is better suited in terms of memory usage and CPU utilization, as package has lesser overhead of resource utilization due to its simplicity of design and native implementation of Julia programming language. This editor is fresh approach for the Julia users which introduces robust Julia terminal editor.

## Tool Availability

The proposed tool is hosted in GitHub repository *([https://github.com/PCTejasvi/Minimalistic-Terminal-Editor-for-Julia-MinTEJ-](https://github.com/PCTejasvi/Minimalistic-Terminal-Editor-for-Julia-MinTEJ-))*. The repository will be made accessible, once corresponding arXiv preprint receives its DOI, thereby facilitating reproducibility and further experimentation.

## APPENDIX

### A. Julia files and description

Table 1. Detailed package organization for different modes in MinTEJ

| | Mode | List of Julia files and their functionalities |
|---|---|---|
| 1. | Editor mode | **edm.jl**: Main function in the editor modes control writing the files and includes all the files below-mentioned.<br>**d.jl**: It provides the functionality of delete the content in the file.<br>**cp.jl**: Provides the functionality to copy the content of file.<br>**bs.jl**: Add the blank line in the editing file based on user request. |
| | | **cm.jl**: Add the comment to the lines based on user input.<br>**uncm.jl**: This functionality provides the uncommenting the lines.<br>**rd_lines.jl**: Provides the functionality to read the files and displays it as output on consoler.<br>**copy_file.jl:** Copy the content of file from one file to other<br>**find.jl:** Search option<br>**undo.jl and redo.jl:** Undoing and redoing the changes made while editing the file |
| **2.** | **File management mode** | **navigation.jl**: This is main file which includes the file management operations and is controlled directly by user options. |
| **3.** | **Execution** | **minirepl.jl:** This file consists of code for direct file execution and REPL based interactive execution. |
| **4.** | **Debug mode** | **db.jl**: This file consists of macros for debugging the Julia code.<br>**dbg_fcn.jl**: This is main function which call the **db.jl** and includes the common function such REPL based debugging in interactive mode. |
| **5.** | **Syntax mode** | **syntax.jl:** consist of code database that user can readily add and use to refer the syntax of Julia commands. |
| **6.** | **Integrated mode module** | All Julia code mentioned above added to module and controlling function **MinTEJ()** which consist of integration of all codes to serve as minimalistic terminal editor which can be exported and readily used for the various mode |

## B. Test code: Memory loading

```
using Dates
using Printf
# Function to simulate memory allocation
function simulate_memory_load(n::Int)
    data = Vector{Vector{Float64}}()
    for i in 1:n
        push!(data, rand(10^6))  # Allocate ~8MB per iteration
        sleep(0.5)              # Simulate gradual load
        @printf("[%s] Iteration %d: Allocated %.2f MB\n", Dates.format(now(),
"HH:MM:SS"), i, Base.summarysize(data) / 1024^2)
    end
    return data
end
# Run the simulation
simulate_memory_load(500)
```